\documentclass[12pt,twoside,a4paper]{article}
\usepackage{amsmath,amssymb,latexsym,theorem,natbib,epsfig,color,subfigure,footmisc,bbm}
\usepackage{multirow,graphicx,array,rotating,url}  
\usepackage{epstopdf,afterpage,wasysym}
\usepackage{verbatim}
\usepackage[normalem]{ulem}
\def\crps{\mathop{\hbox{\rm CRPS}}}
\def\crpss{\mathop{\hbox{\rm CRPSS}}}

\numberwithin{equation}{section}

\title{Clustering-based spatial interpolation of parametric post-processing models}

\author{{S\'andor Baran}$^{1,*}$ and {M\'aria Lakatos}$^{1,2}$  \vspace*{0.5cm}\\
{\small $^1$Faculty of Informatics, University of Debrecen, Hungary}\\
{\small $^2$Doctoral School of Informatics, University of Debrecen, Hungary}
}  

\date{}

\begin{document}

\maketitle

\footnotetext[1]{Corresponding author: \url{baran.sandor@inf.unideb.hu}}
\begin{abstract}
Since the start of the operational use of ensemble prediction systems, ensemble-based probabilistic forecasting has become the most advanced approach in weather prediction. However, despite the persistent development of the last three decades, ensemble forecasts still often suffer from the lack of calibration and might exhibit systematic bias, which calls for some form of statistical post-processing. Nowadays, one can choose from a large variety of post-processing approaches, where parametric methods provide full predictive distributions of the investigated weather quantity. Parameter estimation in these models is based on training data consisting of past forecast-observation pairs, thus post-processed forecasts are usually available only at those locations where training data are accessible. We propose a general clustering-based interpolation technique of extending calibrated predictive distributions from observation stations to any location in the ensemble domain where there are ensemble forecasts at hand. Focusing on the ensemble model output statistics (EMOS) post-processing technique, in a case study based on wind speed ensemble forecasts of the European Centre for Medium-Range Weather Forecasts, we demonstrate the predictive performance of various versions of the suggested method and show its superiority over the regionally estimated and interpolated EMOS models and the raw ensemble forecasts as well.

\end{abstract}

\section{Introduction}
\label{sec1}

From the initiation of operational implementation of ensemble prediction systems (EPSs) in late 1992, both at the European Centre for Medium-Range Weather Forecasts (ECMWF) and the National Centers for Environmental Prediction \citep{b18a}, ensemble-based probabilistic forecasting became the most advanced approach in weather prediction. A forecast ensemble is obtained from different runs of numerical weather prediction (NWP) models, usually with perturbed initial conditions and/or stochastic physics parametrizations. However, even nowadays, when most major weather centres operate EPSs, ensemble forecasts still often suffer from the lack of calibration and might exhibit systematic bias \citep[see e.g.][]{bhtp05,ecmwfEval21}, which deficiencies can be corrected by some form of post-processing \citep{b18b}.

In the last two decades, a large variety of post-processing approaches has been developed for a wide range of weather quantities or even for their combinations; for an overview, we refer to \citet{ppb,vbd21}. Among them, parametric methods such as ensemble model output statistics \citep[EMOS;][]{grwg05}, Bayesian model averaging \citep[BMA;][]{rgbp05}, or the state-of-the-art machine learning-based distributional regression network \citep[DRN;][]{rl18} -- which is the extension of the EMOS -- provide full predictive distributions of the investigated weather quantity.

In contrast, non-parametric approaches are distribution-free and, for instance, can approximate the forecast distribution by estimating its quantiles \citep[various forms of quantile regression; see e.g.][]{fh07,brem19,brem20}, adjust forecasts to match the climatological distribution of observations \citep[quantile mapping;][]{hs18}, or improve the raw ensemble using linear regression \citep[member-by-member post-processing;][]{vsv15}. 

All of the above methods require training data consisting of past forecast-observation pairs, and calibrated forecasts are usually calculated only for those locations where training data are accessible. Since ensemble forecasts are issued on grid points of a given resolution, they can be bilinearly interpolated to any location in the ensemble domain, whereas observations are available mainly at synoptic observation (SYNOP) stations, which are unevenly scattered around the globe. Naturally, gridded reanalyses having the same spatio-temporal scale as the forecasts, thereby enabling interpolation, can also be considered as verification data; however, as argued by \citet{frg19}, post-processing is more efficient ``when trained and verified against station observations''. 

The aim of the present study is to investigate how station-based post-processing can be extended to locations where no observation data are available. We are focusing on EMOS modelling of wind speed ensemble forecasts; however, the proposed methodology can be easily generalized to any parametric approach and any other weather quantity. The EMOS predictive distribution is a single parametric law with parameters depending on the ensemble, and EMOS models for various weather variables usually differ only in the chosen parametric family and/or in the link functions connecting its parameters to the forecast. Here we consider the truncated normal (TN) EMOS approach of \citet{tg10}; although, several other EMOS models are available for wind speed relying on alternative predictive distributions such as generalized extreme value \citep{lt13}, log-normal \citep{bl15} or truncated generalized extreme value \citep{bszsz21}. In the TN EMOS model training data are required to estimate the coefficients of the affine functions expressing the location and scale of the TN distribution as functions of the ensemble members and their variance, respectively. In the case of regional estimation, when data from all available observation stations are comprised into a single training set resulting in a single collection of EMOS parameters shared by all stations, the interpolation to any location in the ensemble domain is straightforward. Using the obtained common set of parameters one can determine the EMOS predictive distribution at any site where forecasts are available. This particular approach is followed e.g. by \citet{bb24}, where after observation-based regional post-processing of solar irradiance using both EMOS and DRN methods, the forecast skill of their interpolation to additional locations is also investigated. However, in the case of large ensemble domains, there might be significant differences in the climatology of the different observation sites, and the properties of the forecast errors can also differ. Hence, the single set of EMOS parameters provided by the regional approach might be undesirable. In such situations local estimation based only on training data of the station of interest is a reasonable option; however, there is no direct way of interpolating local EMOS models to unobserved locations. 

We propose a semi-local interpolation technique, which is an extension of the clustering-based semi-local approach of \citet{lb17} and generalizes both regional and local parameter estimation methods. In a case study based on ECMWF wind speed forecasts, we compare the predictive performance of this approach to parameter estimation and model interpolation to regional and local EMOS modelling and to the raw forecasts as well. All EMOS models are obtained using data of a randomly selected 75\,\% of the studied SYNOP stations, which models are then interpolated to the remaining 25\,\%.

The presented methodology is essentially simpler than the kriging-based approaches of \citet{sb14} and \citet{sm15}, where using additional covariates, the parameters of locally trained EMOS models are interpolated with the help of an underlying Gaussian random field. Furthermore, it also differs fundamentally from the local BMA of \citet{krbgmg11}, which combines the Mass-Baars interpolation \citep{mbwgs08} with a regionally estimated BMA model.

The remaining part of the paper is organized as follows. In Section \ref{sec2} the ECMWF wind speed data used in our case study is introduced. In Section \ref{sec3} we summarize the truncated normal EMOS model, discuss the model verification procedure, and propose the semi-local interpolation algorithm. The results of our case study are reported in Section \ref{sec4}, which is followed by a discussion in Section \ref{sec5}.

\section{Data}
\label{sec2}
\begin{figure}[t]
   \centering
   \epsfig{file=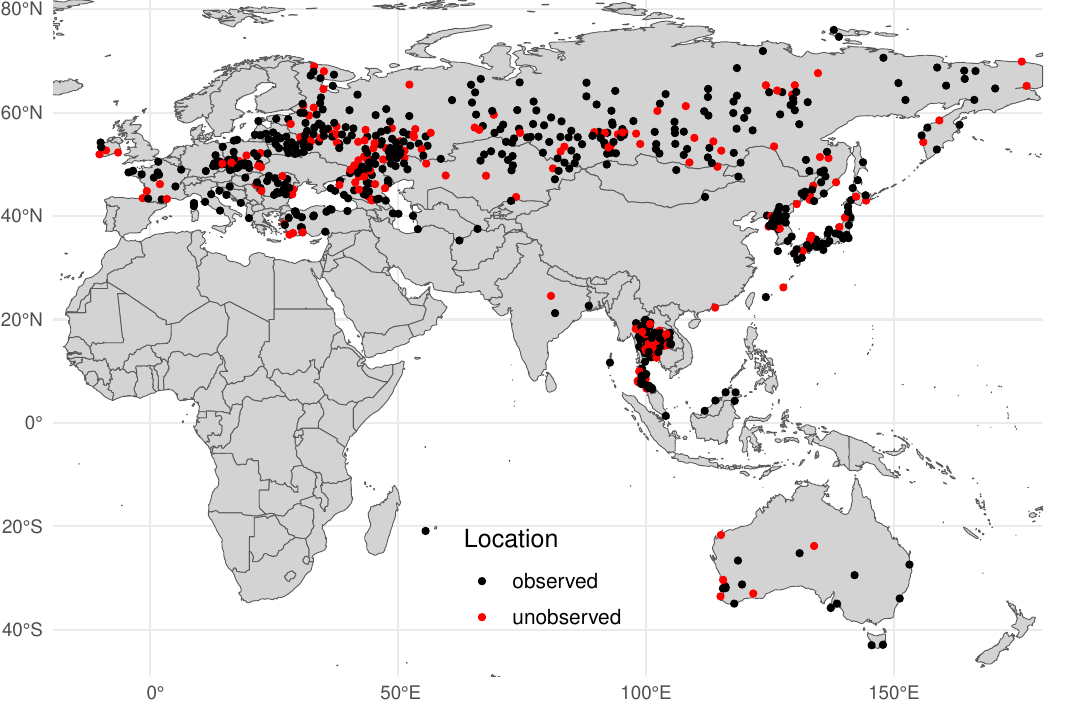, width=.9\textwidth} 
   \caption{Locations of SYNOP stations and the distribution of observed and unobserved locations.}
   \label{fig:map}
 \end{figure}

We consider 52-member ECMWF ensemble forecasts (high-resolution forecast (HRES), control forecast (CTRL), and 50 members (ENS) generated using random perturbations) of 10-m  wind speed for 674 SYNOP stations in Europe, Asia, and Australia for the period between 1 January 2020 and 30 April 2023, together with the corresponding validating observations. All forecasts were initialized at 0000 UTC and we study 10 different lead times ranging from 1 day to 10 days. Randomly selected 75\,\% of the SYNOP stations (506 locations, referred to as {\em observed locations\/}) are used for estimating the parameters of the predictive distributions of wind speed, which models are then interpolated to the remaining 168 locations (referred to as {\em unobserved locations\/}), see Figure \ref{fig:map}. In the dataset at hand around 1.65\,\% of the total forecast cases are incomplete due to missing forecasts or observations.

\section{Post-processing and forecast verification}
\label{sec3}

As mentioned, statistical post-processing of wind speed ensemble forecasts is performed with the help of the TN EMOS approach of \citet{tg10} described in detail  below.

\subsection{Truncated normal EMOS model}
\label{subs3.1}

In what follows, denote by \ $f_1,f_2,\ldots ,f_{52}$ \ the 52-member ECMWF 10-m wind speed forecast of a given forecast horizon for a given location and time point, where \ $f_1=f_{\text{HRES}}$ \ and  \ $f_2=f_{\text{CTRL}}$ \ are the high-resolution and the control forecasts, respectively, while \ $f_3,f_4, \ldots ,f_{52}$ \ stand for the 50 ensemble members obtained using perturbed initial conditions. These members, referred to also as \ $f_{\text{ENS},1}, f_{\text{ENS},2}, \ldots , f_{\text{ENS},50}$, \ can be treated as statistically indistinguishable, hence exchangeable. Furthermore denote by \ $\overline f_{\text{ENS}}$ \ the mean, and by \ $S^2_{\text{ENS}}$ \ the variance of these 50 exchangeable ensemble members, that is
\begin{equation*}
  \overline f_{\text{ENS}}:=\frac 1{50}\sum_{k=1}^{50}f_{\text{ENS},k} \qquad
  \text{and} \qquad S^2_{\text{ENS}}:=\frac 1{49}\sum_{k=1}^{50} \big(f_{\text{ENS},k} - \overline f_{\text{ENS}}\big)^2.
\end{equation*}

Following \citet{gneiting14}, the EMOS predictive distribution of 10-m wind speed based on the 52-member ECMWF forecast ensemble is truncated normal \ ${\mathcal N_0}\big(\mu,\sigma^2\big)$ \ with location \ $\mu$ \ and scale \ $\sigma>0$ \ left-truncated at zero, where
\begin{equation}
  \label{eq:TNlink}
  \mu = a_0 + a_1^2f_{\text{HRES}} +  a_2^2f_{\text{CTRL}} + a_3^2 \overline f_{\text{ENS}} \qquad \text{and} \qquad \sigma^2=b_0^2 + b_1^2S^2_{\text{ENS}}.
\end{equation}
Model parameters \ $a_0,a_1,a_2,a_3,b_0,b_1\in {\mathbb R}$ \ are estimated with the help of training data comprising past forecast-observation pairs. In line with the optimum score principle of \citet{gr07}, we minimize the mean value of a proper verification score (see Section \ref{subs3.2}) over the training data. Note that the predictive performance of post-processed forecasts highly depends on both the chosen verification metric and the spatial and temporal composition of the training data (see Section \ref{subs3.3}).

\subsection{Verification scores}
\label{subs3.2}
In the evaluation of the predictive performance of probabilistic forecasts, \citet{gbr07} suggest the principle of ``maximizing the sharpness of the predictive distributions subject to calibration''. Sharpness is the property of the predictive distribution only and measures its concentration. For instance, for a given \ $\alpha \in ]0,1[$, \ one can investigate the average width of the \ $(1-\alpha)\times 100\%$ \ central prediction intervals, that is intervals between the lower and upper \ $\alpha/2$ \ quantiles of the predictive distribution. The smaller this average width, the more concentrated the predictive distribution. In contrast, for a calibrated probabilistic forecast predictive distributions and corresponding observations should be statistically consistent, so calibration is their joint property. Considering again the above central prediction intervals, for a calibrated forecast their coverage (proportion of observations located in the corresponding central prediction intervals) should be around \ $(1-\alpha)\times 100\%$. \ Note that by choosing \ $\alpha$ \ to match the nominal coverage of \ $96.23\%$ \ of the 52-member ECMWF ensemble, one can provide a fair comparison with the raw forecasts.

However, these two goals can be addressed simultaneously with the aid of proper scoring rules quantifying predictive performance. Here we focus on the  continuous ranked probability score \citep[CRPS;][Section 9.5.1]{w19}, which for a (predictive) cumulative distribution function (CDF)  \ $F$ \ and a real-valued observation \ $x$ \ is defined as
\begin{equation}
    \label{eq:CRPSdef}
\crps(F,x) := \int_{-\infty}^{\infty}\big[F(y)-{\mathbb I}_{\{y\geq x\}}\big]^2{\mathrm d}y
\end{equation}
with \ ${\mathbb I}_H$ \ denoting the indicator function of a set \ $H$. \ Note that smaller CRPS means better fit and in the estimation of the EMOS model parameters (see Section \ref{subs3.1}) the mean of this particular score is minimized.

In order to highlight better the differences between the investigated probabilistic forecasts, we also report the continuous ranked probability skill scores \citep[CRPSS; see e.g.][]{gr07}. The CRPSS quantifies the improvement in the mean score value \ $\overline{\crps}$ \ of a given forecast over the verification period with respect to the corresponding mean \ $\overline{\crps}_{ref}$ \ of a reference method and defined as
\begin{equation*}
  \crpss := 1 - \frac{\overline{\crps}}{\overline{\crps}_{ref}}.
\end{equation*}
We note that skill scores are positively oriented, that is the larger the better.

A further tool for assessing the calibration of ensemble forecasts is the verification rank histogram  \citep[see e.g.][Section 9.7.1]{w19}. It displays the histogram of ranks of the verifying observation with respect to the corresponding forecast ensemble. The ranks of a properly calibrated 52-member ensemble follow a discrete uniform distribution on integers \ $\{1,2,\ldots,53\}$, \ and the shape of the rank histogram can refer to bias in probabilistic forecasts and/or their under- or overdispersive character. In the case of probabilistic forecasts given in the form of predictive distributions, a similar role is played by the probability integral transform \citep[PIT;][Section 9.5.4]{w19} histogram. PIT is the predictive CDF evaluated at the verifying observation, which for a well calibrated probabilistic forecast follows a standard uniform law.

Furthermore, the accuracy of forecast medians and means, serving as point predictions, is evaluated with the help of the mean absolute error (MAE) and the root mean squared error (RMSE), respectively. This choice is based on the fact that the median optimizes the former, whereas the mean minimizes the latter \citep{gneiting11}.

Finally, in order to assess the statistical significance of the differences between the verification measures, we report confidence intervals for the skill scores and score differences. These intervals are based on 2000 block bootstrap samples calculated using the stationary bootstrap scheme with average block length determined according to \citet{pr94}.

\subsection{Clustering-based model estimation and interpolation}
\label{subs3.3}

As mentioned in Section \ref{subs3.1}, parameters of the EMOS model \eqref{eq:TNlink} are estimated using training data, where both temporal and spatial composition might affect the predictive performance of post-processed forecasts. Here we consider rolling training periods, which are based on ensemble forecasts and corresponding observations of the preceding \ $n$ \ calendar days. This popular approach allows quick adaptation, for instance, to model changes or seasonal variations; however, one can also consider monthly, seasonal, or yearly windows \citep[see e.g.][]{hhp16,jmg23}, whereas machine learning-based techniques requiring large training datasets usually rely on static training data \citep[see e.g.][]{gzshf21,sl22}.

Concerning the spatial selection of training data, a basic approach is local modelling \citep{tg10}. In this case, the predictive distribution for a given location is estimated using only past forecast-observation pairs of that particular location. In general, locally trained models result in the best forecast skill, provided the training period is long enough to circumvent numerical problems in modelling. For an overview of optimal training period lengths for various weather quantities, we refer to \citet{hspbh14}. A disadvantage of local modelling is that it does not allow a direct interpolation of predictive distributions to locations, where only forecasts are available without the corresponding validating observations. On the contrary, regional models are based on training data of the whole ensemble domain, thus providing a single set of EMOS parameters for all investigated locations. Using these parameters one can extend the predictive distribution to all sites of the studied region where there are ensemble forecasts at hand. Compared to local modelling, the regional approach requires shorter training periods. Nevertheless, regional models are not suitable for large heterogeneous domains and usually underperform their local counterparts.

To combine the beneficial properties of local and regional modelling, \citet{lb17} propose a clustering-based semi-local approach. Based on individual feature vectors assigned to each observation station, which represent climatology and performance of the ensemble forecast during the training period, locations are first arranged into clusters using $k$-means clustering, then within each cluster, a regional estimation is performed.
This selection of features, discussed in details below and referred to as {\em observation-based,\/} appeared to be successful in several case studies involving different weather quantities such as temperature \citep{blszbb19}, heat indices \citep{bbpbb20} or precipitation \citep{szgb23}.

\subsubsection*{Observation-based quantile features} 
Following the notations of \citet{lb17}, let \ $\widehat F_{i,n}$ \ and \ $\widehat G_{i,n}^e$ \ denote the empirical CDFs of station climatology (observed wind speed) and forecast errors of the mean \ $\overline f_{\text{ENS}}$ \ of the exchangeable ensemble members at location \ $i$ \ over the rolling training period of length \ $n$, \ respectively. The $N$-dimensional feature vector assigned to station \ $i$ \ is defined as the concatenation of equidistant \ $\frac 1{N_1+1}, \frac 2{N_1+1}, \ldots ,\frac {N_1}{N_1+1}$ \ quantiles of  \ $\widehat F_{i,n}$ \ and  equidistant \ $\frac 1{N_2+1}, \frac 2{N_2+1}, \ldots ,\frac {N_2}{N_2+1}$ \ quantiles of  \ $\widehat G_{i,n}^3$, \ where \ $N_1=\lfloor \frac N2 \rfloor$ \ and \ $N_2=N-N_1$. \

\medskip
From the perspective of model interpolation, the disadvantage of the above feature set is that it uses observations, hence it is not straightforward how to assign an unobserved location to one of the existing clusters. A possible solution is to consider features that are based merely on the ensemble forecasts. This approach will be called {\em forecast-based\/} and we consider two different setups.

\subsubsection*{Forecast-based quantile features}
This feature set is the forecast-based analogue of the previous, observation-based construction. Denote by \  $\widehat H_{i,n}^{(1)}$, \  $\widehat H_{i,n}^{(2)}$, \  $\widehat H_{i,n}^{(3)}$ \ and \  $\widehat H_{i,n}^{(4)}$ \ the empirical CDFs of  high-resolution forecasts \ $f_{\text{HRES}}$, \ control forecasts  \ $f_{\text{CTRL}}$ \ and  mean \ $\overline f_{\text{ENS}}$ \ and standard deviation \ $S_{\text{ENS}}$ \ of the 50 exchangeable ensemble members, respectively, at location \ $i$ \  over the training period of length \ $n$. \ Now the \ $N$ \ dimensional feature vector for station \ $i$ \ comprises \ the concatenation of equidistant quantiles of  \ $\widehat H_{i,n}^{(j)}$ \ at levels \ $\frac 1{N_j+1}, \frac 2{N_j+1}, \ldots ,\frac {N_j}{N_j+1}, \ j=1,2,3,4,$ \ where $N_1=N_2=N_3=\lfloor \frac N4\rfloor$ \ and \ $N_4 = N-N_1-N_2-N_3$.

\subsubsection*{Forecasts as features}
Instead of equidistant quantiles, for each location \ $i$ \ and training period of length \ $n$, \ consider simply the $4n$-dimensional concatenation of training vectors of \ $f_{\text{HRES}},  \ f_{\text{CTRL}}, \ \overline f_{\text{ENS}}$ \ and \ $S_{\text{ENS}}$. \ Naturally, this simple approach might result in rather high-dimensional features, which increases the computational cost of $k$-means clustering. However, the additional cost is usually less than the cost of computing the equidistant quantiles comprising a forecast-based quantile feature.

\bigskip
As forecast-based features are available also at those sites, where there are no observations at hand (unobserved locations), we propose the following interpolation algorithm.
\begin{itemize}
   \setlength\itemsep{1mm}
\item[]{\em Step 1.\/} Using either observation-based features, or one of the forecast-based approaches, estimate clustering-based semi-local EMOS models for the observed locations.
\item[] {\em Step 2.\/} Choose one of the forecast-based feature selection methods (not necessarily the same as in Step 1). To each observed location assign the corresponding feature vector and calculate the cluster means.
\item[] {\em Step 3.\/} Calculate the same forecast-based features as in Step 2 for unobserved locations.
\item[] {\em Step 4.\/} Assign each unobserved location to the closest cluster in terms of the Euclidean distance of the feature vector of the given location to the cluster mean.
\item[] {\em Step 5.\/} For each unobserved location use the EMOS parameters of the assigned cluster to calculate the corresponding predictive distribution.
\end{itemize}
Note that by setting the number of clusters to be equal to the number of observed locations one gets back the local approach, whereas the use of a single cluster results in regional estimation. In the former case, the above interpolation algorithm assigns each unobserved location to the closest observed one in terms of the Euclidean distance of their feature vectors. 

A natural question is whether instead of the above algorithm, interpolation can be performed simply based on the geographical locations. This approach would assign an unobserved location to the closest cluster, for instance, in terms of the mean geographical distance to the cluster members. In particular, in the local case, it would mean that an unobserved location would inherit the EMOS parameters of the geographically closest observed one. However, at least for the dataset at hand, the geographical interpolation consistently underperforms the proposed algorithm in all reported semi-local cases and for all lead times. For local models the situation is slightly different as for short lead times the use of geographical coordinates results in a minor benefit compared to the forecast-based matching; nevertheless, even this local model is behind in skill of the interpolated forecast-based semi-local methods.

\section{Results}
\label{sec4}

\begin{figure}[t]
   \centering
   \epsfig{file= 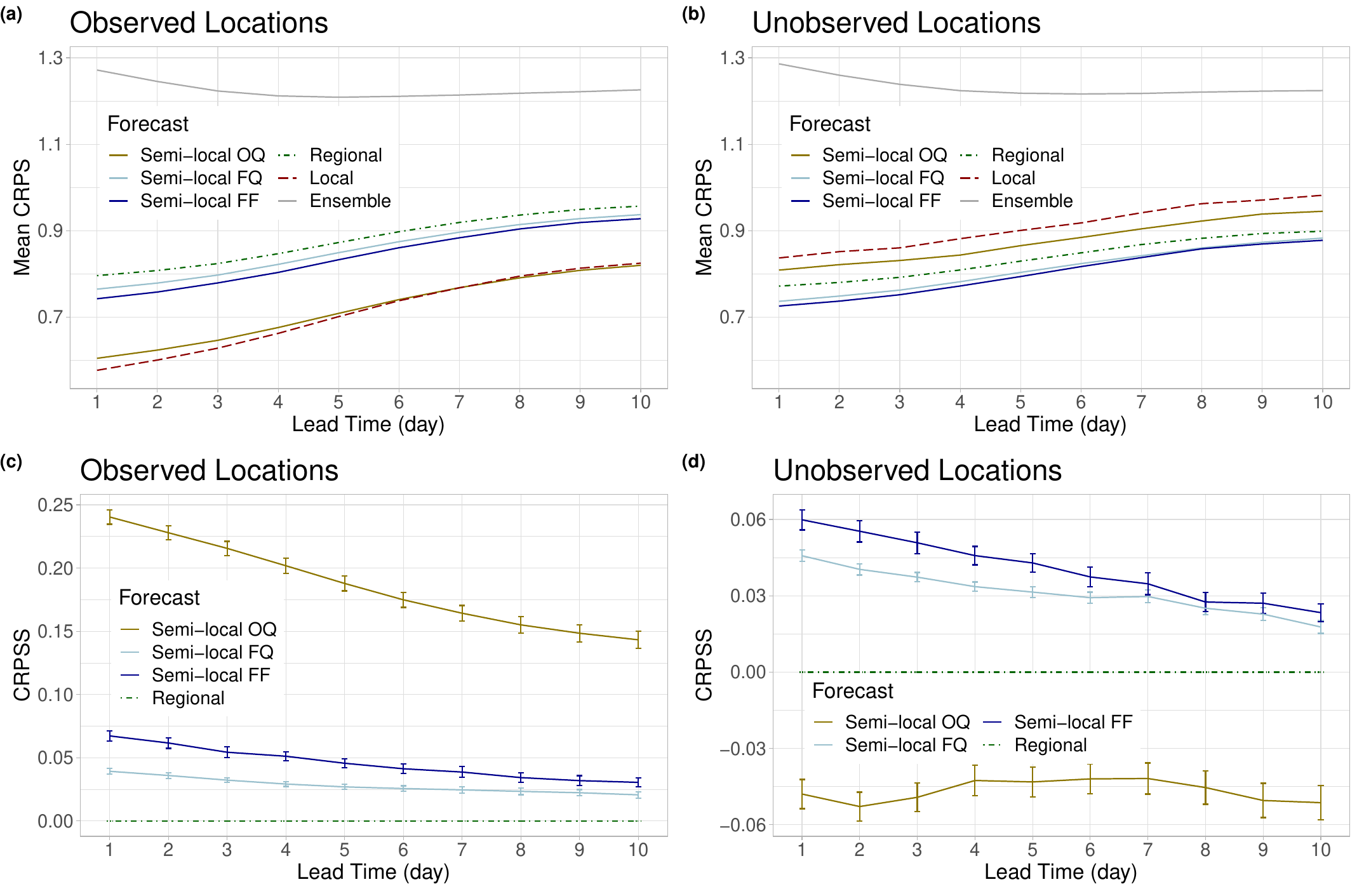, width=\textwidth} 
   \caption{CRPS of post-processed and raw forecasts for observed (a) and unobserved (b) locations, and CRPSS of semi-local models with respect to the regional approach for observed (c) and unobserved (d) locations together with 95\,\% confidence intervals as functions of the lead time.}
   \label{fig:crps_crpss}
 \end{figure}
 
The efficiency of the interpolation algorithm presented in Section \ref{subs3.3} in terms of the predictive performance of post-processed forecasts at unobserved locations is tested with the help of ECMWF wind speed data introduced in Section \ref{sec2}. Based on the three investigated feature sets, we consider three different semi-local EMOS models for the observed locations (Step 1), which will be referred to as {\em semi-local OQ\/} (observation-based quantile features), {\em semi-local FQ\/} (forecast-based quantile features) and {\em semi-local FF\/} (forecasts as features). According to \citet{lb17}, provided one has enough features to represent the empirical CDFs, the size \ $N$ \ of the quantile-based features has only a marginal effect. Hence, in the observation-based case we set \ $N=24$ \ as in \citet{lb17} and \citet{szgb23}, whereas in the forecast-based one, similar to \citet{bbpbb20}, 40-dimensional features are considered. Observed locations are arranged into 20 clusters, and unobserved locations are assigned to clusters (Steps 2 and 3) utilizing forecasts as features, since this approach results in more skillful models than the use of forecast-based quantile features. As a reference post-processing approach, we consider the regional TN EMOS model based only on the observed locations and interpolated to the unobserved sites; however, we also report the scores of the local TN EMOS model and its interpolation with the help of forecasts as features, and the raw ECMWF ensemble forecasts. All EMOS models are trained using a 60-day rolling training window, and forecast-observation pairs of the period 1 May 2020 -- 30 April 2023 are considered for validation purposes. Note that semi-local models based on 10 and 15 clusters and training period lengths of 30 and 90 days were also tested. In general, the predictive performance of the EMOS models exhibits only a very slight dependence on the training period length with 60 days being the optimal one, and fewer clusters result in slightly worse forecast skill for all investigated forecast-based semi-local models and training windows.

\begin{figure}[t]
   \centering
   \epsfig{file= 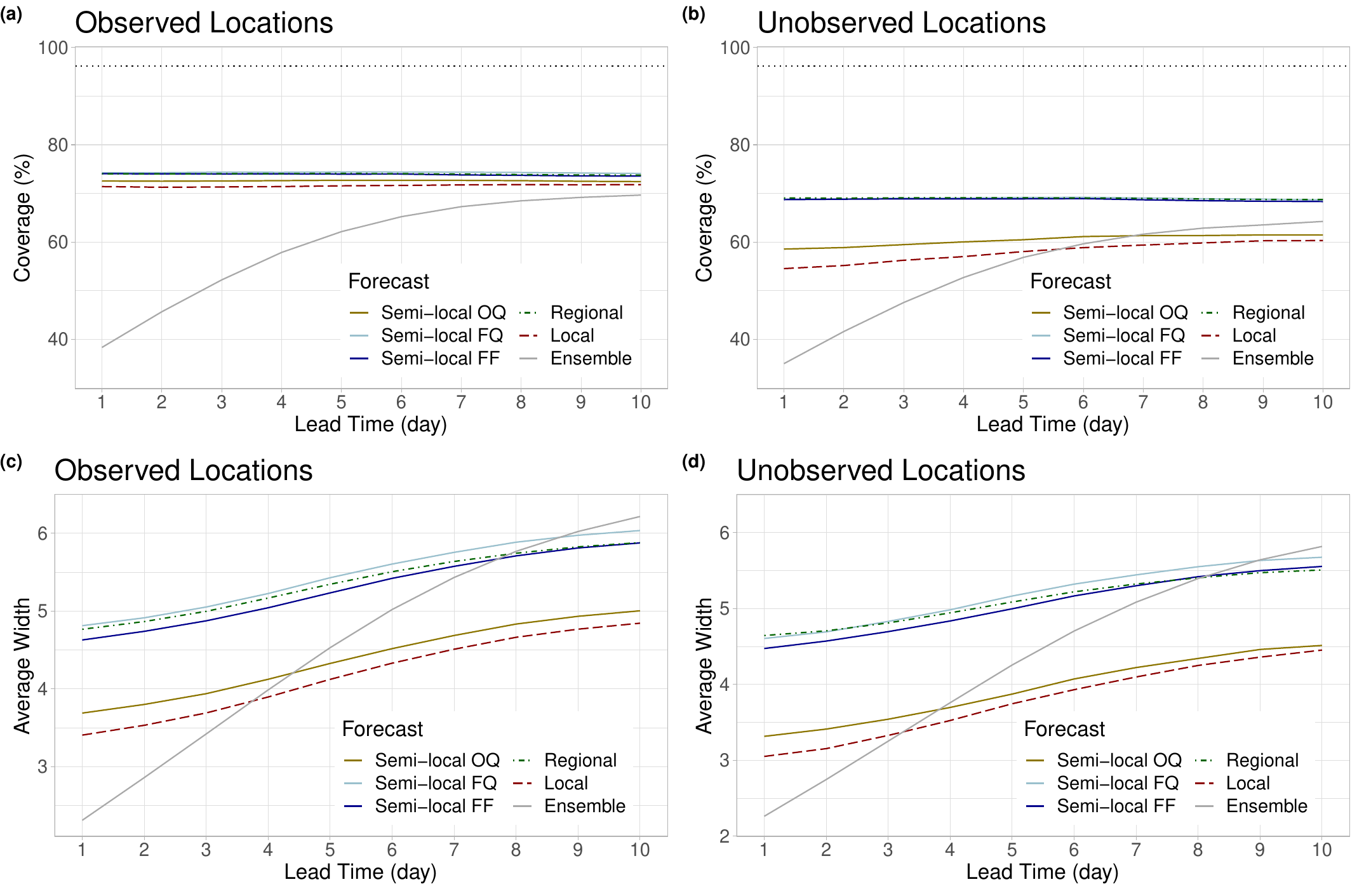, width=\textwidth} 
   \caption{Coverage (a,b) and average width (c,d) of nominal 96.23\,\% central prediction intervals of post-processed and raw forecasts for observed (a,c) and unobserved (b,d) locations functions of the lead time. In panels (a) and (b) the ideal coverage is indicated by the horizontal dotted line.}
   \label{fig:cov_aw}
 \end{figure}

 Panels (a) and (b) of Figure \ref{fig:crps_crpss} show the mean CRPS of post-processed and raw forecasts for observed and unobserved locations, respectively. As expected, all calibrated forecasts outperform the raw ensemble by a wide margin, and the skill of the various EMOS models degrades with the increase of the forecast horizon. Note that the different shape of the mean CRPS of the raw forecasts is due to the representativeness error in the verification, which behaviour has been observed in other case studies as well \citep[see e.g.][]{bszsz21}, and can be partially corrected by the perturbed-ensemble approach of \citet{bbhwhr20}. At the observed locations the local TN EMOS model exhibits the best overall predictive performance, closely followed by the observation-based semi-local approach (OQ) of \citet{lb17}, which after day 7 results in the lowest mean CRPS. Note that for a 30-day training period the semi-local OQ EMOS model outperforms the local one after day 3 (not shown), whereas among EMOS models based on a 90-day training window, the local one is the most skillful for all investigated forecast horizons (not shown). This observation is in line with the arguments in the discussion of \citet{frg19}, that the longer the lead time, the more training data is required for efficient calibration. Furthermore, according to the skill scores of Figure \ref{fig:crps_crpss}c, at observed locations all semi-local EMOS models significantly outperform the regional one with the observation-based (OQ) approach being far the best, and the difference between the two forecast-based methods is also significant for all lead times. As panels (b) and (d) of Figure \ref{fig:crps_crpss} indicate, the lack of observations drastically changes the model ranking. Local model results in the highest mean CRPS and while the performance of the two forecast-based semi-local EMOS models is similar at observed and unobserved locations, the observation-based semi-local EMOS is significantly behind its competitors. In general, at unobserved locations the skill of best performing forecast-based semi-local FF EMOS approach is 2.3 -- 6\,\% ahead of the performance of the regional TN EMOS model, whereas at observed sites its advantage is in the 3 -- 6.7\,\% interval.

\begin{figure}[ht!]
   \centering
   \epsfig{file= 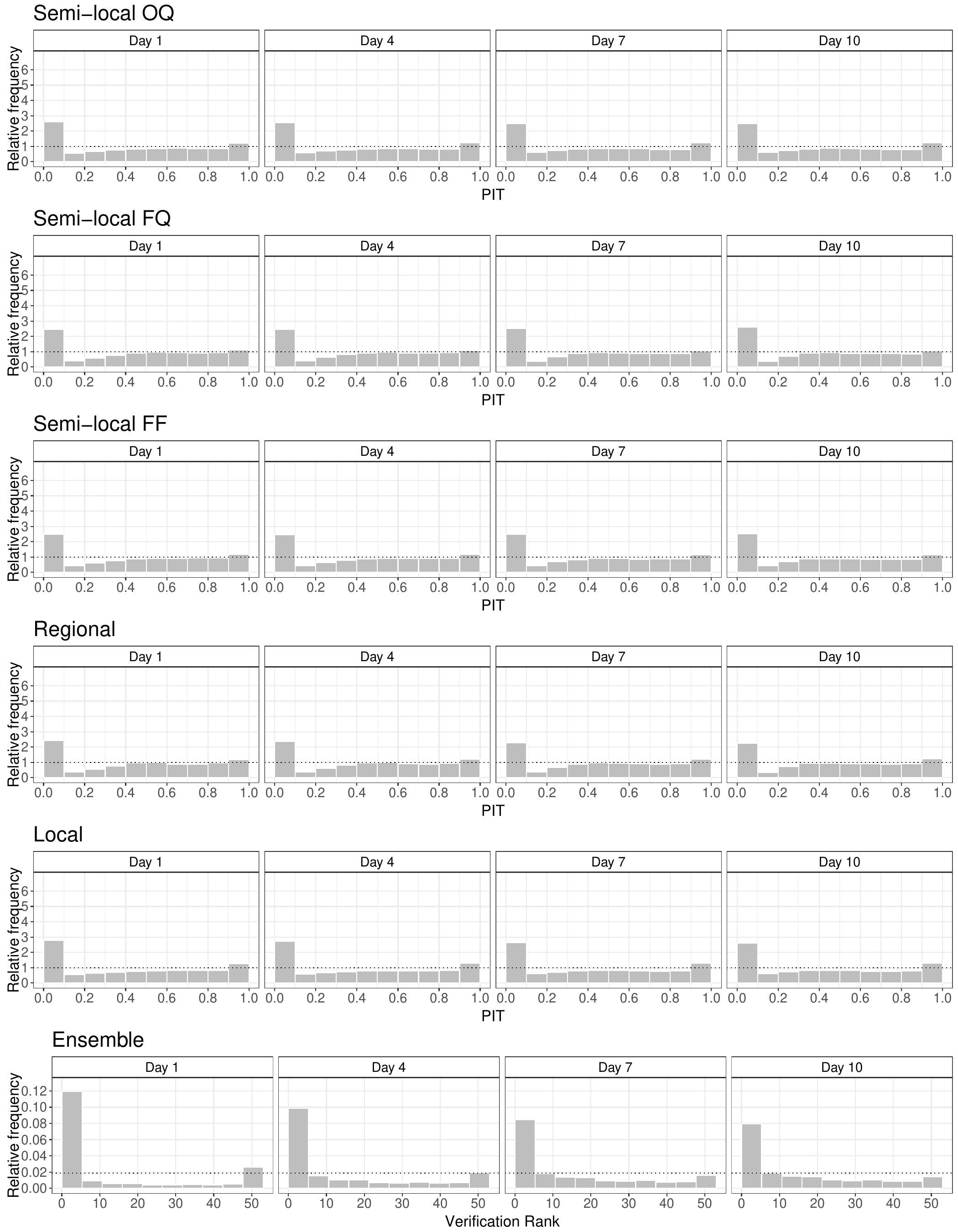, width=\textwidth} 
   \caption{PIT histograms of post-processed and verification rank histograms of raw ensemble forecasts of wind speed at observed locations for lead times 1, 4, 7 and 10 days.}
   \label{fig:pitO}
 \end{figure}

 \begin{figure}[ht!]
   \centering
   \epsfig{file= 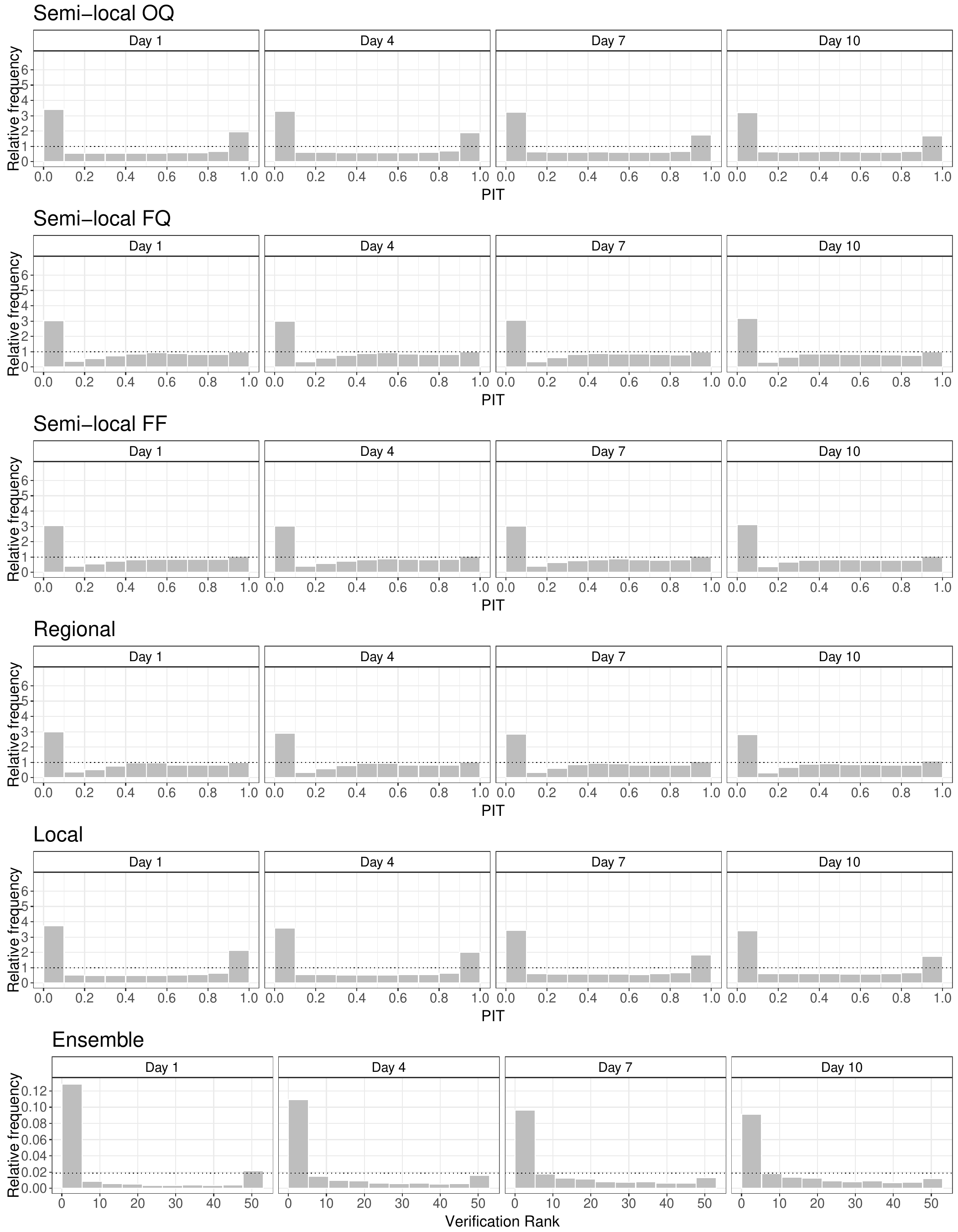, width=\textwidth} 
   \caption{PIT histograms of post-processed and verification rank histograms of raw ensemble forecasts of wind speed at unobserved locations for lead times 1, 4, 7 and 10 days.}
   \label{fig:pitU}
 \end{figure}

Next, as displayed in Figure \ref{fig:cov_aw}a, at observed locations all post-processing approaches substantially improve the coverage of the raw ECMWF wind speed forecasts, which ranges from 38.32\,\% to 69.65\,\% and increases with the increase of the forecast horizon. Regional and forecast-based semi-local EMOS models (FQ and FF) result in the highest and almost identical coverage values with a slightly decreasing trend; however, their coverage of around 74\,\% is still far from the nominal 96.23\,\% level. Moreover, from day 8 the regional and semi-local FF models result in sharper central prediction intervals than the raw ensemble (Figure \ref{fig:cov_aw}c). Note that the coverages of observation-based semi-local (OQ) and local approaches resulting in far the sharpest predictive distributions are just slightly behind the other three models. According to panels (b) and (d) of Figure \ref{fig:cov_aw}, at unobserved locations the behaviour of raw forecasts and interpolated regional and forecast-based semi-local EMOS models are rather similar to the observed case; nonetheless, there is a slight systematic decrease of around 4 percentage points in all coverage values. In contrast, the coverages of the local and observation-based semi-local EMOS models are substantially lower; from days 6 and 7, respectively, they are even below the coverage of the raw ensemble.

\begin{figure}[t]
   \centering
   \epsfig{file= 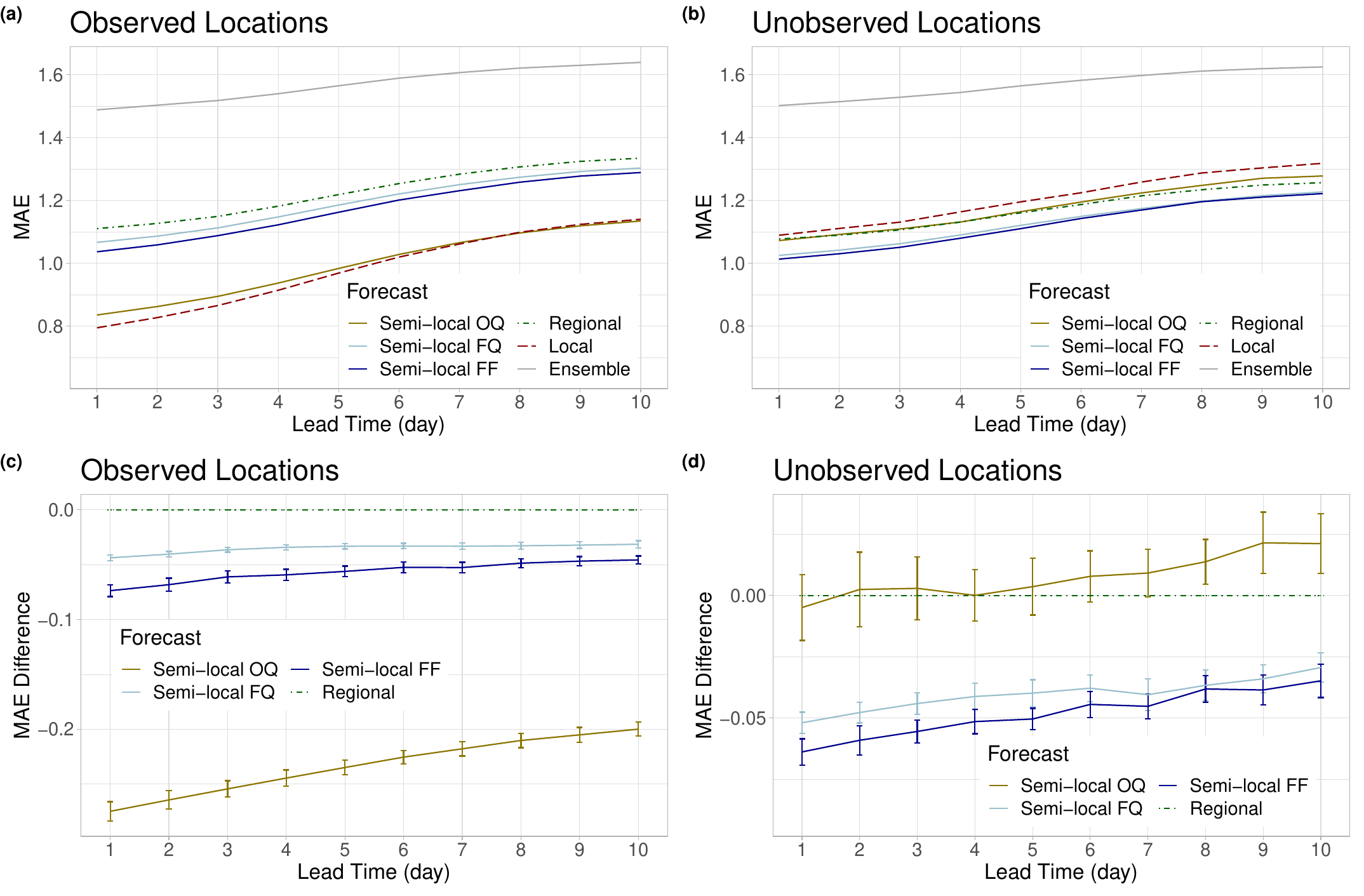, width=\textwidth} 
   \caption{MAE of the median forecasts for observed (a) and unobserved (b) locations, and difference in MAE of semi-local models from the regional approach for observed (c) and unobserved (d) locations together with 95\,\% confidence intervals as functions of the lead time.}
   \label{fig:mae_maed}
 \end{figure}

The verification rank and PIT histograms of Figures \ref{fig:pitO} and \ref{fig:pitU} provide further evidence of the improved calibration of post-processed forecasts both at observed and unobserved locations. Raw forecasts are highly underdispersive (U-shaped) with a strong positive bias at both sets of sites; nevertheless, these deficiencies are less and less pronounced with the increase of the forecast horizon. Post-processing considerably decreases underdispersion and to a certain extent also the bias. Regional and forecast-based semi-local methods result in rather similar PIT histograms both at observed and unobserved locations; however, in the latter case, the bias correction is slightly less efficient. In contrast, at unobserved locations, the PIT histograms of local and observation-based semi-local (OQ) EMOS models are far more underdispersive than their counterparts at the observed sites.

Finally, panels (a) and (c) of Figure \ref{fig:mae_maed} showing the MAE of median forecasts and the difference in MAE of semi-local EMOS approaches from the regional EMOS tell the same story as the corresponding panels of Figure \ref{fig:crps_crpss}. At observed locations local and observation-based semi-local EMOS calibration is the most beneficial and the MAE values of regional and semi-local models differ significantly; however, the maximum difference in MAE between the best and worst performing EMOS model is around 0.3 m/s. In contrast, while the behaviour and ranking of regional and forecast-based semi-local models, similar to the mean CRPS, remains the same at unobserved sites (panels (b) and (d) of Figure \ref{fig:mae_maed}), up to day 7 there is no significant difference in terms of MAE between the regional and the observation-based semi-local EMOS model. Note that the behaviour and ranking of the different probabilistic predictions in terms of the RMSE of the mean forecasts is rather similar to the case of the mean CRPS, hence it is not shown.

Based on the above analysis one can conclude that clustering-based interpolation of semi-local EMOS models considering forecasts as features is superior to the regional approach to parameter estimation and results in more skillful predictive distributions at locations where no observations are available.

\section{Discussion}
\label{sec5}

We propose a novel algorithm for interpolating parametric predictive distributions to locations where only ensemble forecasts are available without corresponding validating observations, hence direct post-processing is not possible. At observed locations, where training data of forecast-observation pairs are accessible, model parameters for each time point are estimated using the clustering-based semi-local approach of \citet{lb17}; however, besides observation-based $k$-means clustering of stations in terms of their similarity in station climatology and forecast error of the ensemble mean, merely forecast-based grouping of stations is also investigated. After having predictive distributions at observed stations, unobserved locations inherit model parameters of the closest cluster in terms of similarity in the corresponding ensemble forecasts to the cluster mean. This approach generalizes regional modelling, where a single set of parameters is valid for the whole ensemble domain, and hence can be used directly for interpolation.

In contrast to the spatial interpolation methods proposed either by \citet{krbgmg11} or by \citet{sb14} and \citet{sm15}, the presented methodology does not need the estimation of an underlying Gaussian field or any additional station-specific information; it relies only on the data required for EMOS modelling. Hence, it can be easily generalized to any parametric post-processing method including the DRN models as well. 

As expected, at observed sites, local and observation-based semi-local approaches to parameter estimation result in the best predictive performance, followed by forecast-based and regional models. However, this ranking drastically changes when unobserved locations are considered. Forecast-based semi-local approaches consistently outperform their observation-based counterpart and local modelling in terms of all investigated verification measures. Compared with the regional approach, the best performing semi-local model based on forecasts as features results in sharper central prediction intervals subject to the same coverage level and a significant improvement in terms of the mean CRPS, the MAE of the median, and RMSE of the mean forecasts. 

A possible direction of future research is the study of interpolation of station-based post-processing to gridded dual-resolution forecasts \citep[see e.g.][]{szgb23}. The actuality of the problem comes from the introduction of the new CY48R1 model cycle of the ECMWF Integrated Forecast System \citep{ecmwf23} in June 2023, issuing 51 medium-range forecasts at TCo1279 resolution ($\approx$ 9 km) and 101 extended-range predictions at TCo319 resolution  ($\approx$ 32 km).

Another potential course of further studies is to investigate the forecast skill of multivariate post-processing methods such as the ensemble copula coupling \citep{stg13} or the Gaussian copula approach \citep{mlt13} interpolated to unobserved (possibly gridded) locations. For an overview of multivariate approaches, we refer the reader to \citet{sm18}, whereas \citet{lbm20} and \citet{llhb23} provide detailed comparisons of these methods using simulated data and real ensemble forecasts, respectively.

\section*{Acknowledgements}  The authors gratefully acknowledge the support of the  National Research, Development and Innovation Office under Grant No. K142849 and the S\&T cooperation program 2021-1.2.4-T\'ET-2021-00020.
Finally, they are indebted to Ivana Aleksovska for providing the ECMWF wind speed data.

\end{document}